\documentclass{vgtc}                          

\ifpdf
  \pdfoutput=1\relax                   
  \usepackage{graphicx}                
  \DeclareGraphicsExtensions{.pdf,.png,.jpg,.jpeg} 
\else
  \usepackage{graphicx}                
  \DeclareGraphicsExtensions{.eps}     
\fi%

\graphicspath{{figures/}{pictures/}{images/}{./}} 
\usepackage[shortcuts]{extdash}
\usepackage{microtype}                 
\PassOptionsToPackage{warn}{textcomp}  
\usepackage{textcomp}                  
\usepackage{mathptmx}                  
\usepackage{times}                     
\usepackage{tabularx}
\usepackage{cite}                      
\usepackage{tabu}                      
\usepackage{booktabs}                  
\usepackage{caption}
\usepackage{subcaption}
\usepackage{enumitem}
\usepackage{setspace}

\onlineid{0}

\vgtccategory{Research}

\title{Cross-Reality for Extending the Metaverse: \\Designing Hyper-Connected Immersive Environments with XRI}

\author{Jie Guan \thanks{e-mail: jie.guan@ocadu.ca} %
\and Alexis Morris\thanks{e-mail:amorris@ocadu.ca}
\and Jay Irizawa  \thanks{e-mail:jirizawa@ocadu.ca}}
\affiliation{\scriptsize Adaptive Context Environments (ACE) Lab \\ OCAD University}   

\abstract{The Metaverse comprises technologies to enable virtual twins of the real world, via mixed reality, internet of things, and others. As it matures unique challenges arise such as a lack of strong connections between virtual and physical worlds. This work presents design frameworks for cross-reality hybrid spaces. Contributions include: i) clarifying the metaverse "disconnect", ii) extended metaverse design frameworks, iii) prototypes, and iv) discussions toward new metaverse smart environments.

} 

\CCScatlist{
 Virtual Reality, Mixed Reality, Augmented Reality, Extended Reality, Internet of Things, Human Computer Interaction}

\begin{document}
\begin{titlepage}

     \vspace{1cm}
        Full Citation: J. Guan, A. Morris and J. Irizawa, "Cross-Reality for Extending the Metaverse: Designing Hyper-Connected Immersive Environments with XRI," 2023 IEEE Conference on Virtual Reality and 3D User Interfaces Abstracts and Workshops (VRW), ShangHai, China, 2023, pp. 305-311, DOI: 10.1109/VRW58643.2023.00071.

       \vspace*{1cm}

       \copyright2023 IEEE. Personal use of this material is permitted.  Permission from IEEE must be obtained for all other uses, in any current or future media, including reprinting/republishing this material for advertising or promotional purposes, creating new collective works, for resale or redistribution to servers or lists, or reuse of any copyrighted component of this work in other works.

       \vspace{1.5cm}
  
\end{titlepage}


\firstsection{Introduction}

\maketitle




The metaverse refers to ``a virtual environment blending physical and digital, facilitated by the convergence between the Internet and web technologies, and Extended Reality'' \cite{Lee2021a}. This blended environment promises numerous benefits for society, which can be environmental, social, economic, and even psycho-physiological; as this technology has broad human-factor impacts on society such as the physical, psychological, social, organizational, and political dimensions \cite{vicente2013human}. For example the metaverse can impact the environment when largely adopted, by reducing transportation costs and resources within the environment, while simultaneously enhancing people’s efficiency when working and studying remotely in dynamic shared virtual spaces
. 

Moreover, the time spent in metaverse environments that are more dynamic, immersive, and creative has the potential to extend these human factors; from the world of physical limitations to the more unlimited virtual environment. This is currently being adopted by large-scale industry developments and the metaverse has become one of the most influential trends in the marketplace. Various industries, such as health-care\cite{thomason2021metahealth}, marketing and advertising \cite{swilley2016moving}, travel and tourism \cite{gursoy2022metaverse}, education and training\cite{Collins2008} are adapting their services to provide metaverse-enabled networked platforms. 


Today's metaverse applications remain limited, however, as they are primarily online social experiences composed of virtual environments, virtual objects, and virtual avatars. This introduces a set of metaverse challenges for humans-in-the-loop. One of these challenges is the metaverse "disconnect" problem, as in \cite{guan2022extendingThesis}.
Diverse metaverse platforms and virtual spaces host activities with standalone computers, mobile devices, and head-mounted display VR devices. Most of them bring humans into complete virtual spaces, thereby focusing on the virtual content leading to the metaverse dynamics being independent of - or disconnected from - the physical space. 
This disconnect presents a challenge to the adoption of the metaverse, as there is a need to allow the human-in-the-loop to be more aware of real-world environmental contexts as they simultaneously become more immersed in virtual and mixed reality environments.




\subsection{Extending the Metaverse with Cross-Reality XR-IoT (XRI)}

An extended metaverse offers the potential to inhabit two spaces concurrently. One space is the physical world, and the other is the metaverse, which is constructed of computational graphics and simulated environments. 
Living and working in immersive hybrid virtual and physical worlds with smart devices (i.e., head-mounted displays) may become as commonplace as the mobile phone today. Within these speculative futures parameters, this work explores an extended metaverse framework to enhance embodiment, interaction, and agency \cite{Holz2011} that could provide a seamless interface to the physical world.

To explore designs for this new interface across both physical and virtual sides of the metaverse, requires a hybrid approach that encompasses the ubiquitous computing and immersive interface domains, within what is considered as "Cross-Reality," or "X-Reality (XR)," i.e., the ubiquitous mixed reality environment when  combined with context sensing \cite{paradiso2009guest} \cite{Morris2021}. 
This work achieves this by extending traditional IoT interfaces with XR-IoT (XRI) interfaces, toward hybrid virtual and physical objects.

A hybrid approach exploring both physical and virtual attributes of the metaverse is needed in developing a new and cohesive interface. Previous works have presented hybrid mixed-reality and internet-of-things frameworks (known as XR-IoT, or XRI) \cite{Morris2021}\cite{Tsang2021}, and more recent work has presented early prototypes \cite{Guan2022IEEEVR}, \cite{guan2022extendingThesis} including an approach to strengthen the connection between physical spaces and metaverse environments (such as an  XRI Lamp controller and an XRI Ambient Lighting system connecting real physical lamps to their virtual counterparts \cite{Guan2022IEEEVR}). The current paper furthers these research contributions for extending the metaverse, based on \cite{guan2022extendingThesis}.

The project explores an extended metaverse framework for applying the metaverse layer in the physical space(s) to increase dynamic inter-connections of humans, agents, and the environment through MR and the IoT. The extended metaverse framework focuses on improving interaction, embodiment, and agency \cite{Holz2011} in the human-in-the-loop MR space.
The contributions of this include: i) an exploration of the metaverse disconnect problem and background literature, ii) an architectural framework for extending the metaverse, iii) design prototypes using the framework, (one for a hyper-connected space for time-awareness while in the metaverse; and the other for enabling users to transition between real and virtual environments where the environment helps to detect physical objects), and iv) a discussion-based evaluation of these approaches. The problems of the current metaverse platforms have yet to address these models of connection with the physical space in depth, and this work has the potential to 
elucidate the disconnect problem and to lay the groundwork for future exploration toward seamless immersive engagements in metaverse applications.

\section{Background and Related Works}
To discuss the concept of extended metaverse frameworks and agents, this section provides the background knowledge of the metaverse, related theory of XR and the IoT, and introduction of the context awareness along with generative and procedural content concepts. The background of the metaverse addresses the history, current state of technological development, and future possibilities. 

XR is the method to embody the virtual and physical objects and environment while the IoT is considered to be for virtual and physical communication. Context awareness is introduced to capture and sense information in the physical environment, and procedural content is the method and rules to embody and enhance dynamic virtual objects.

\subsection{The Background of Metaverse}
As mentioned by Dionisio et al.\cite{Dionisio2013}, the metaverse is a portmanteau that combines with the prefix “meta,” which means “beyond,” and the suffix “verse,” which is shorthand for “universe.” It represents a universe beyond the space we live in physically. Specifically, it is a computer-generated environment that simulates the world and distinguishes it from the "metaphysical" or "spiritual" concepts. 
The concept of the metaverse, originally from the fiction novel Snow Crash, was developed by Neal Stephenson in 1992 \cite{Joshua2017}. In the novel, the metaverse was portrayed as a virtual world with humans interacting with intelligence agencies and each other as avatars in that space. It is a scenario similar to that in Ready Player One by Ernest Cline, where users can adopt any role or play as any character-type in a completely virtual world \cite{Ai2021}
The metaverse has various definitions (see table \ref{Comparisonofpapersaboutthemetaverse}) because it is a state-of-the-art term that is being continually explored. Dionisio et al.\cite{Dionisio2013} presented that the metaverse is constructed by multiple individual virtual worlds, defined as a fully immersive, three-dimensional digital environment that reflects the totality of shared online space. Lee et al. \cite{Lee2021a} considered the metaverse as a virtual environment constructed by the internet, web technologies, and extended reality (XR) toward hybrid physical and virtual space.

\begin{table*}[t]
\tiny
  \centering
  \begin{tabular}{ |m{11em}|m{30em}|m{23em}| } 
  \hline
    Literature & Metaverse Definition & Purpose of Article \\
    \hline
    Metaverse Theory (2021) \cite{Ai2021} & The metaverse is a sci-fi world from Snow Crash that enables people to do similar activities as in Ready Player One \cite{Cline2011}. & To guide the development of future blockchain games within the context of the metaverse. \\
    \hline
    3D Virtual Worlds and the Metaverse (2013)  \cite{Dionisio2013} & The metaverse is a fully immersive, three- dimensional digital environment constructed by the interconnection of individual virtual worlds. & To present a survey and history of the metaverse. \\
    \hline
    Escaping the Gilded Cage (2004) \cite{Ondrejka2004} & The metaverse is an online environment that is real for users, which can interact with each other similar to the real world. & To show the economic and legal decisions that maximize the power of player creativity for the metaverse. \\
    \hline
    Higher Education in the metaverse CHRIS (2008)  \cite{Collins2008} & The metaverse is beyond the vision of Stephenson’s immersive 3D virtual world, to include the aspects of physical world objects that are virtually-enhanced physical and physically persistent virtual space. & To address the
problem of higher education and use the metaverse platform for teaching. \\
    \hline
    
    A content service deployment plan for metaverse museum (2017) \cite{Choi2017} & The metaverse includes AR and simulation as one of the fundamental axes (lines), and another is internal and external elements. The key elements of the metaverse include AR, virtual worlds, lifelogging, and mirror worlds. & To provide a metaverse exhibition experience service that allows users to journey back and forth between real and virtual spaces. \\
    \hline
    
    Distribute the Metaverse (2018) \cite{Ryskeldiev2018} & The metaverse is a persistent and constant collection of MR space mapped into different geospatial locations. &To decrease the computational costs for mobile MR applications and expand available interactive spaces. \\
    \hline
    
    All One Needs to Know about the Metaverse (2021) \cite{Lee2021} & The metaverse is a virtual space that blends the digital and physical world with the internet, web, and extended reality (XR) technologies. & To provide a survey to offer a comprehensive view of the metaverse that includes multiple underlying technologies and the resulting social ecosystem. \\
    \hline

  \end{tabular}
  \caption{Comparison of selected papers about the metaverse, as in \cite{guan2022extendingThesis}.}
  \label{Comparisonofpapersaboutthemetaverse}
\end{table*}

There are many applications and platforms for the metaverse today. Understanding their features and limitations will provide a better direction for addressing metaverse disconnect. Digital ownership is a big problem in the metaverse, and blockchain is applied in some platforms to deal with it, such as Decentraland\footnote{https://decentraland.org/}, Cryptovoxels\footnote{https://www.cryptovoxels.com/}, and Somnium Space\footnote{https://somniumspace.com/}. Decentraland attempts to be a virtual reality platform based on the Ethereum blockchain, where users own the creation’s property entirely when they purchase the “land.” The land is a non-fungible token (NFT), a digital asset stored as an Ethereum smart contract that users can create \cite{Ordano2017}. Cryptovoxels and Somnium Space also have a similar approach to defining the ownership of digital assets. Although Decentraland’s VR feature is still in development, Cryptovoxels’s "Origin City" is available to be visited through VR in other popular VR worlds like VRChat, NeosVR, and Substrata. Meanwhile, Somnium Space allows for VR support via common SteamVR supported VR headsets. 

\subsection{Extended Reality and Internet-of-Things}
\textbf{XR-IoT Theory:} As examined in works like \cite{Morris2021} and \cite{Tsang2021},  designing a more connected metaverse requires the merger of technology paradigms of XR and IoT -- to connect with the physical environment while extending into the virtual environment. Augmented Reality provides this with an interactive medium of overlaid virtual objects anchored to the real environment, while IoT refers to the networking of physical objects with computing devices for sensing and communication, as examined in \cite{Jo2019survey}. This hybridization, as in \cite{Tsang2021} is here referred to as XR-IoT (XRI), which represents the combination of XR-based IoT systems as well as IoT-based XR systems. XRI engages in immersive, information-rich, multi-user, and agent-driven systems  \cite{Morris2021}. The combination of these technologies has the potential to bring a closer connection between humans and their environmental objects, as well as each other, and future hybrid XRI applications are being developed for applied situations, like education, cyber security, and marketing\cite{AndradeBastos2019}.


\textbf{XR-IoT Applications:} A selection of these XRI applications is described below, highlighting those projects which embed IoT systems into mixed reality projects, and those that apply XR features in IoT system designs. These metaverse systems merge multiple technologies, as well as interaction techniques, as in \cite{zhao2022metaverse}, in order to create cross-reality experiences. Much of these works are recent and gaining momentum as both paradigms attain maturity and adoption. For instance, the series of XR-IoT projects presented IoT Avatars and workstation prototypes. The IoT Avatar was started with a simple proof of concept that embodies the MR representation Avatar for a plant and provides buttons to control physical servo motors and LEDs of an IoT device through a mobile phone\cite{shao2019iot}. This has been extended to explore MR frameworks for IoT with head-mounted displays for immersion and expressiveness \cite{Guan2020}\cite{Morris2020IOTAvatar}. The work used video-passthrough MR and collected real-time context of a plant (i.e., its lighting intensity, soil moisture, and the number of people nearby to attend to the plant), which was used to project emotional states of a virtual plant avatar via fuzzy logic. Based on the avatar, XRI workstation\cite{Morris2021} extended the avatar by expanding environmental objects (virtual desk) into the scenario to explore a workstation use case. Similarly, “Digi-log” \cite{jo2019Digi-log+} provided an augmented reality shopping scenario using IoT-enabled products within a seamless and scalable AR service. The project presented data visualization based on object position, and mechanisms for access, control, interaction, and content interoperability. A further example, Seiger et al.\cite{seiger2021holoflows} presented HoloFlows, a new MR interaction method for end-users to manage standard IoT devices without coding. Together these early explorations have set the stage toward the extended metaverse themes of this research. 

\section{An Architecture for Extended Metaverse Agents} \label{ArchitextureSection}



\begin{figure*}[tbh]
 \centering 
 \includegraphics[width=0.9\linewidth]{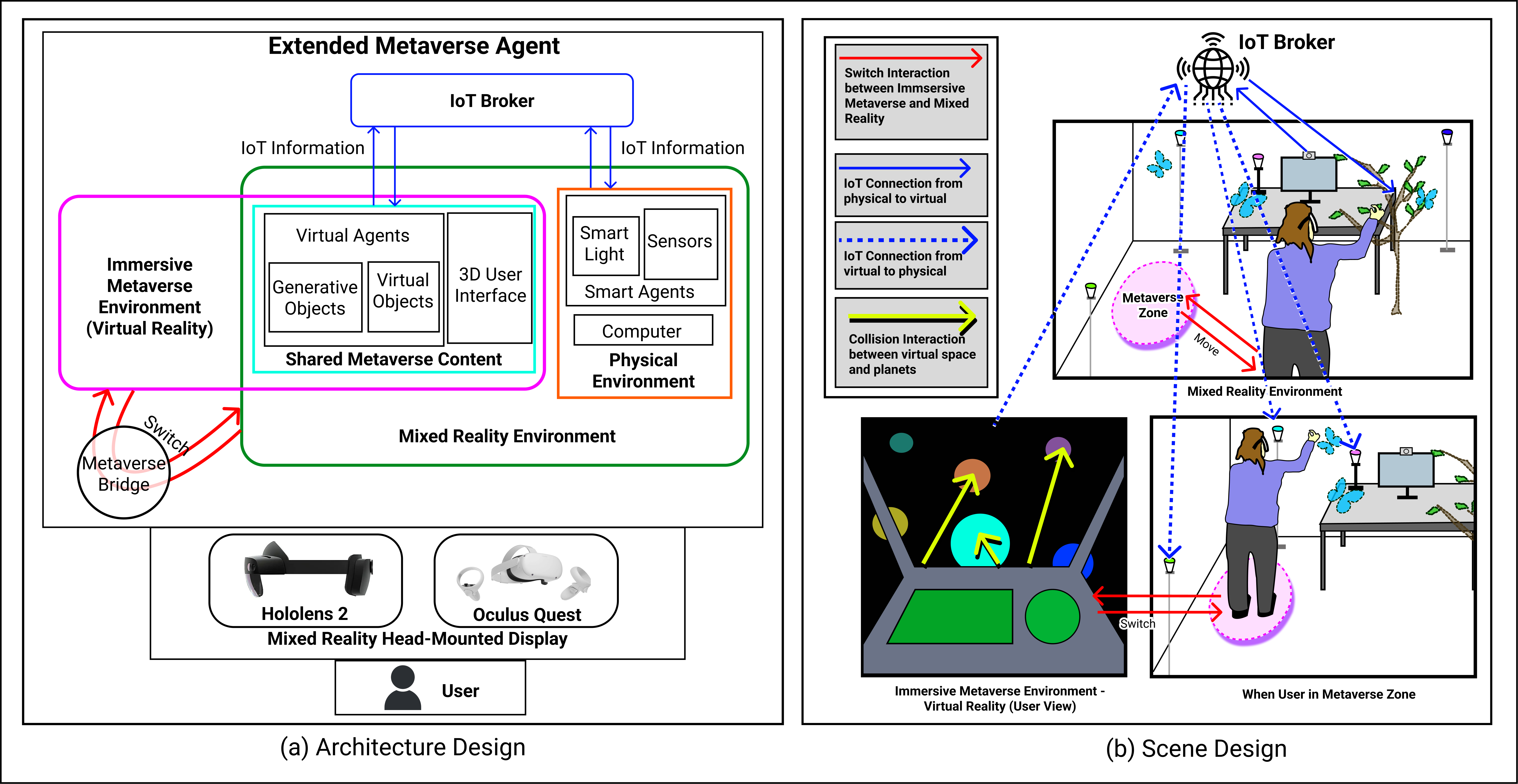}
 \caption{(a) Conceptual Architecture for extended metaverse framework systems that strongly connect the real and the virtual (i.e., the entire RV spectrum \cite{Milgram2011} \cite{guan2022extendingThesis}). A metaverse bridge enables switching between immersive VR and XR environments while an IoT broker enables communication and control of IoT physical and virtual objects.  (b) Reference illustration of a scene design reflecting the architecture and details related to user position, environment lighting, and immersive content during an example metaverse interaction.}
 \label{Architecture}
\end{figure*}


Figure \ref{Architecture} presents a scene design and architecture for XRI systems that are oriented toward addressing the metaverse disconnect challenge through a convergence of prior research on the Metaverse, XRI, context awareness, Mixed Reality Agents, presence and generative design \cite{Guan2022IEEEVR}\cite{Morris2021}\cite{Tsang2021}.

\subsection{Architecture Component Design}
The architecture in (Figure \ref{Architecture}(a)) outlines the design components needed for providing users with an immersive extended metaverse. This includes the user's physical environment, the user's mixed-reality environment (and HMD displays), and an immersive metaverse environment (virtual reality content), and a bridge between these environments. The system enables users to transition between these hybrid spaces, as shown by the red arrows. The blue arrows highlight the bi-directional information communication between the shared metaverse content and the physical environment. These components are described below: 




\textbf{Physical Environment} refers to the real space humans live in every day, and all the objects there could be interacted with physically. In the extended metaverse agent design, it has smart agents performing as actuators (smart light as an example) and sensors (Arduino sensors and computer visions), and a computer to operate the system (to perform the operation). These agents are IoT-enabled with informational communication (shown as the blue arrows) between physical and virtual agents (named as shared metaverse content), that could be affected by each other. 

\textbf{Shared Metaverse Content} refers to the virtual elements (including virtual agents and 3D user interface) that could be access both in mixed reality and the immersive metaverse environment. The virtual agent is the virtual representation (for full or partial embodiment) of the smart agents \footnote{Note that in this work embodiment of IoT objects is considered as the expression of the physical object via its virtual or hybrid virtual-physical counterpart.}, that may also perform tasks. The generative objects of the metaverse content are non-static objects with fixed embodiment (3D models), that also change and grow dynamically. In the design of this architecture, the generative objects are influenced in their behavior physical context information. Additionally, the elements of 3D user interfaces and widgets is represented, accounting for the new forms of user interaction becoming common for virtual content interfaces that differs from the traditional 2D interface on screen-based devices.

\textbf{Mixed Reality Environment} refers to the hybrid virtual and physical environment, where users see the computer-generated graphic and real space together (through optical or video pass-through)\cite{Milgram2011}. It includes the physical environment and the shared metaverse content which represented the virtual elements. 

\textbf{Immersive Metaverse Environment} refers to a completely synthetic world that could immerse the user in totally \cite{Milgram2011}, with constructing by multiple virtual worlds in fully immersive environment\cite{Dionisio2013}.

\textbf{IoT Broker:}The smart agents with sensors in physical environment has the ability to capture context information that could affect and control the shared metaverse content (such as affecting the generative objects, and controlling the behaviours of virtual agetns) through IoT broker. Likewise, the smart light showed in the component represented the physical actuators, which could affected and controlled by the information from shared metaverse content (such as the interacting with the 3D user interface and virtual objects collision) through IoT Broker. 

\textbf{Metaverse Bridge:} act as the method for user to switch between the mixed reality environment and the immersive metaverse environment. With this bridge, it could enhance the usability of virtual reality applications since it could increase the accessible of physical environment which users rely on.

\subsection{Scenario Design}
Figure \ref{Architecture}(b) presents an overall scenario reflecting the architecture components and details. For the mixed reality environment, the user is wearing a head-mounted display headset and could see both the virtual elements (shared metaverse content) and the physical environment. Users could see a virtual tree, virtual butterflies, and a virtual circle represented the metaverse zone on the space. A webcam is attached to the monitor for capturing the physical context through computer vision, and such information could affect and control the virtual elements through IoT communication indicated as solid blue arrows. A metaverse zone is designed to reflect the metaverse bridge in the architecture design, it allow users to move in and move out to switch between immersive metaverse environment and mixed reality environment, as indicated with red arrows. When in the immersive metaverse environment, the user is able to operate a virtual spaceship and perform collision interaction with the virtual planets (indicated with yellow arrows), and such interaction could control the color of physical ambient lighting as indicated with dashed blue arrows through IoT broker.

\section{Extended Metaverse Design Prototypes:\\ MetaPlant and Meta-RV-Traveller}

\begin{figure}[htb]
 \centering 
 \includegraphics[width=\linewidth]{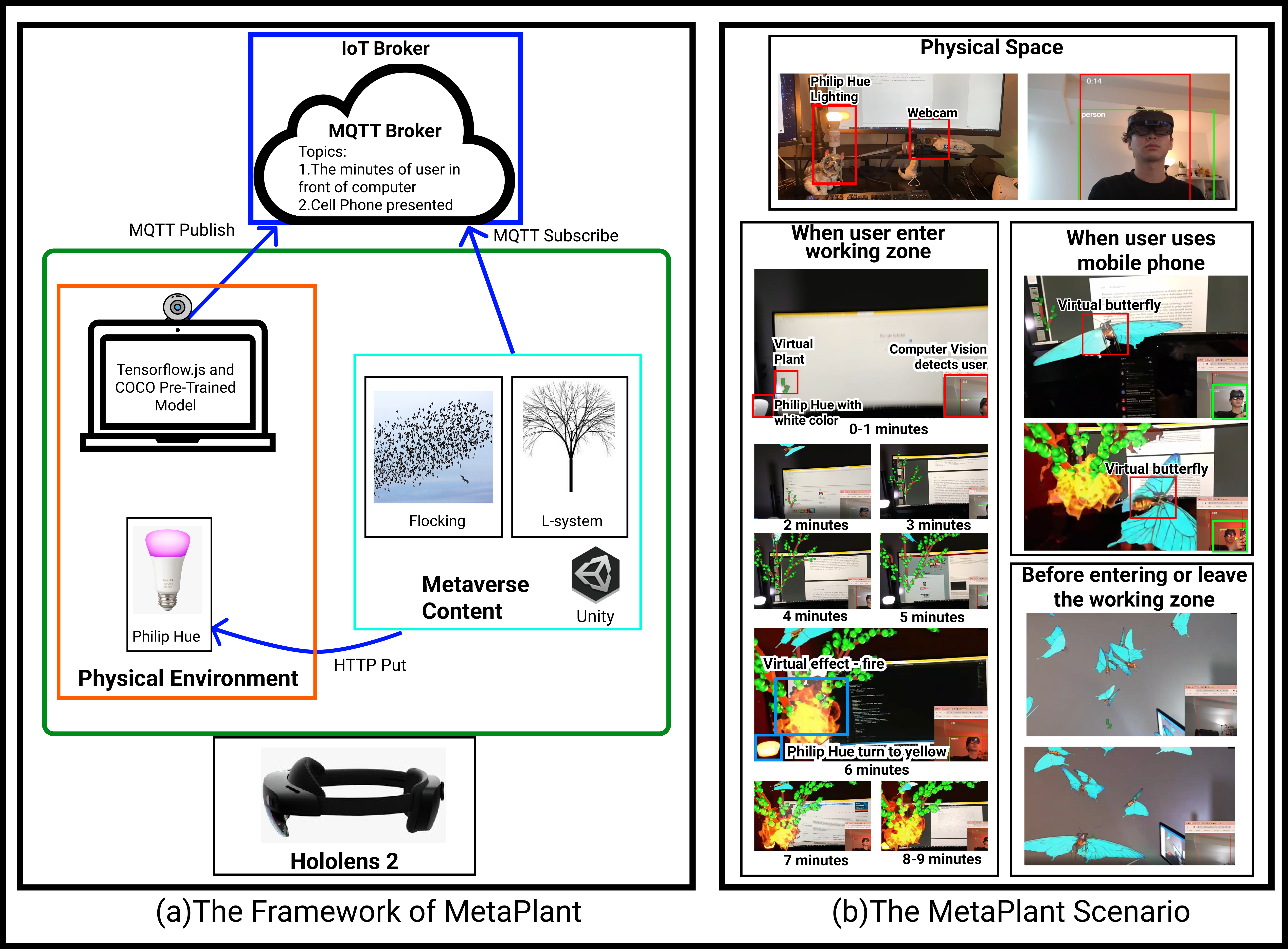}
 \caption{(a) The framework of MetaPlant with the IoT (MQTT and HTTP protocol) connection between physical (computer vision model detect environment context) and virtual (Unity visualization with Hololens 2) environment, based on \cite{guan2022extendingThesis}. (b) The MetaPlant scenario connects user context with mixed reality and the physical environment (ambient lighting). A growing tree represents time-context, and a merger of visual effects, and physical lighting dynamics enrich the environment.}
 \label{MetaPlant}
\end{figure}

\begin{figure*}[tbh]
 \centering 
 \includegraphics[width=0.9\linewidth]{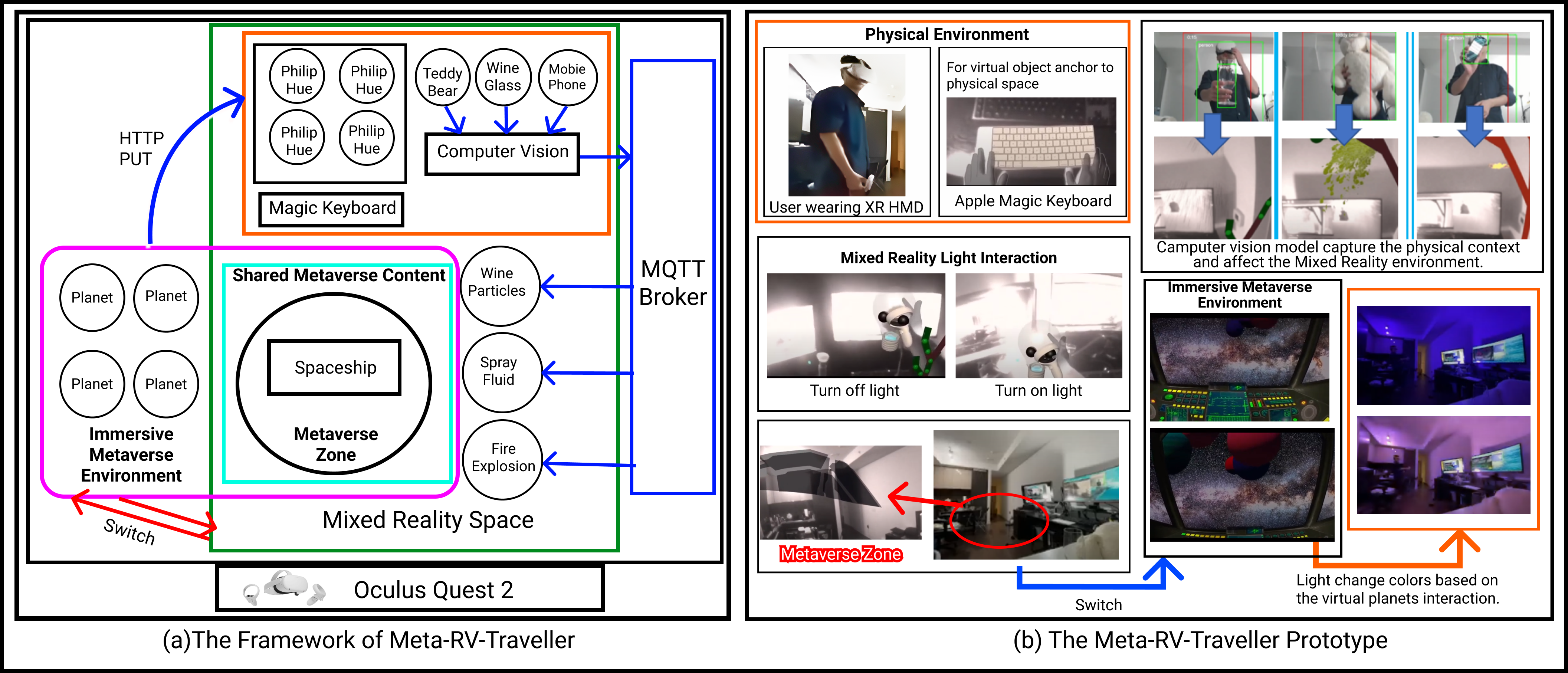}
 \caption{(a) Prototype Framework for Transitioning in a Hyper-Connected Metaverse Environment, as in \cite{guan2022extendingThesis}. (b) The Meta-RV-Traveller prototype explores possible user interactions in the hyper-connected metaverse, as the user transitions between virtual, mixed, and physical environments.}
 \label{Meta-RV-Traveller}
\end{figure*}

The architecture has been instantiated into two prototype scenarios for demonstrating interaction and design within the extended metaverse scenarios. This includes a design exploration of user awareness of time spent in a work context, as in Figure \ref{MetaPlant}; and a similar exploration with a focus on user transitions across immersive contexts, as in Figure \ref{Meta-RV-Traveller}. Each prototype is described in the following subsections below, based on \cite{guan2022extendingThesis} \cite{Guan2022IEEEVR}.

\subsection{MetaPlant: An Extended Metaverse Scenario for Cross-Reality Awareness of Time Context}

The goal of this prototype is to illustrate a more complex extended metaverse agent scenario with a generative object and context awareness approach in a smart space. This prototype (see Figure \ref{MetaPlant}) uses location (computer vision detects a user in front of their desktop) and timing (the length of time a user sits in front of the computer) as context to effect the shape, presentation, and growth of the MetaPlant (a virtual plant in the metaverse) and the color of physical ambient lighting. Also, using a mobile phone will attract a virtual butterfly to fly around the head of the user. This project aims to build an immersive and dynamic “clock” to provide a ten-minute block alarm to prevent users from sitting too long in front of computers and the distraction of virtual butterflies flying around when the user uses a cellphone at work. It is a form of data visualization project that represents the time as the virtual tree grows and turns the minutes to the iteration of the L-system. 

This project provides two-way communication (information connection from virtual to physical and vice versa) between the metaverse and physical space. In addition, it is a proof-of-concept prototype that demonstrates how the generative and procedural virtual object through the L-system and flocking algorithm could enhance the dynamic of the extended metaverse environment. As demonstrated in Figure \ref{MetaPlant}(b), a Philip Hub ambient light is presented as an actuator by changing color in the physical space. At the same time, a webcam can be used as a sensor in a computer vision model through a web browser to detect whether the user is in front of the computer. In this scenario, the user is wearing a HoloLens 2 to enter an extended metaverse space that could see the virtual embodiment of the plant agent and butterfly. The virtual plant has eight states, with iterations growing with L-system based on the time, combined with expressive mixed reality effects, such as fire on the virtual plant (over time) and corresponding change in physical lighting color (after a fixed time-frame, in minutes).



The framework (see Figure \ref{MetaPlant}(a)) of MetaPlant contains an MQTT Broker that addresses the IoT protocol on the cloud that is used to connect the computer vision model and the Unity environment. A computer vision model in Tensorflow.js \footnote{https://www.tensorflow.org/js} is available via a browser, and detects whether the human is visible in the space over time and whether they are using a mobile phone (in this case this is whether or not the phone is visible), which then publishes this context information to the MQTT Broker with ``Minutes'' and ``Cell Phone Presented'' topics. Unity is the visualization and interaction engine used to subscribe to these topics and use these data to generate a virtual plant with L-system and operate the butterfly movement. Regarding the physical implementation, a Philip Hue was considered in Prototype 2 to change a physical agent that presented the two-way communication between the metaverse and physical space.


The system starts with object detection when the class detected is equal to “person,” and the timer starts. Likewise, when the label is equal to “cell phone,” a timer of the cell phone being in proximity begins. Communication of this data is via a private MQTT broker powered by Mosquitto\footnote{https://mosquitto.org/} for IoT publish and subscribe messaging. Additionally for the selected IoT devices, the physical light (Philip Hue), is controlled by sending HTTP Put requests with JSON data.


In terms of the visualization of the virtual content in HoloLens 2 with Unity, the L-system generates the virtual plant with iterations and the flocking behaviors of the butterfly. The L-system starts with pre-set parameters to indicate the growing shape of the virtual tree in different iterations. The butterfly’s movement is controlled by a conventional flocking engine (Cloud Fine). For butterfly movement the head of the user and the location of the virtual plant allow for queuing behavior settings, allowing them to be dynamic in the user's environment, while switching between these two positions in different conditions.

This prototype uses a simple way for the detection of context, by capturing human and mobile phone present in front of the computer. By combining the color control of the light, as in \cite{Guan2022IEEEVR}, it can perform two-way communication between metaverse content and the physical agent. With the generative object of a virtual plant, the system is considered to enhance the overall dynamics and promote an engaging MR experience (it is noted that evaluating this experience remains for future testing with users). 



\subsection{Meta-RV-Traveller: Cross-Reality Transitions in a Hyper-Connected Metaverse Environment}
Meta-RV-Traveller explores the concept of how users can engage and ``travel'' between real and virtual environments based on the virtuality continuum \cite{Milgram2011} in the metaverse context. 
Figure \ref{Meta-RV-Traveller} presents the prototype framework designed for exploring this transition aspect of the extended metaverse, and the resulting prototype exploration is shown in Figure \ref{Meta-RV-Traveller}(b). This  involved designing a metaverse zone in the physical space by combining previous work \cite{Guan2022IEEEVR} of switching capacity between two spaces. The virtual environment (a spaceship and universe background) would appear immersive when users enter the area. 
A virtual bulb is also a shared object of the metaverse virtual and mixed reality environment. The user could also switch on the physical light by moving in/out in a specific area in the MR space and can bring this object with them into the metaverse (fully virtual) zone where they can use it to control the spaceship movement in the virtual universe. In the virtual environment, the bulb attaches to a virtual joystick to control the movement of the spaceship. Two-way communication from virtual to physical and from physical to virtual is shown by switching on and off the physical light using a virtual bulb as one of the interactions from virtual to physical communication. Additionally, when driving the spaceship in the metaverse, there are many planets moving toward the users. Users can move the spaceship by manipulating the virtual joystick to avoid or collide with the planets. If the spaceship hits a planet, then the physical ambient light in the room will change the color to that of the recently collided planet.

On the other hand, similar to previous prototype, the from physical to virtual interaction uses computer vision to detect physical context. The generative tree, from previous, is also incorporated into this project, plus object detection to affect the virtual presentation. When the computer vision model detects example objects (in this case a wine glass, teddy bear, and phone), the MR environment will relatively display a representative visualization (e.g., flowing wine particles, spray fluid, and fire explosion), thereby enabling the fully immersed user to experience the physical context.

\section{Discussion}

The proof-of-concept prototypes presented in the previous section are designed to be representative and exploratory. They highlight the potential for interaction simultaneously between the metaverse environments, mixed reality, and the physical environment. They explore possible control of devices in one part of the spectrum that influence those in another part of the spectrum. These also bring attention to the need for designs that bring the benefits of virtual and mixed reality to communicate to the user in rich visual dimensions. Likewise, they also show the merger of IoT sensing and control, with computer vision, can be actively applied to mixed reality scenes for environment understanding and presentation of a shared context, which will be essential for a more hyper-connected metaverse. Although this work is not comprehensively evaluated (such as via user study), the frameworks developed are a step toward future research in this direction. However, to show how these prototypes can relate to other work in this area, a subjective rating has been applied, based on a selection of factors (inspired by \cite{Holz2011}\cite{Milgram2011}\cite{Jovanovic2022}\cite{Zeltzer1992}) considered deemed important for such systems.

\begin{table*}[htb]
\tiny
  \centering
  \begin{tabular}{ |m{7em}|m{11em}|m{7em}|m{10em}|m{11em}|m{13em}|m{13em}|} 
  \hline
    Design Factors & Description & Very Low & Low & Medium & High & Very High \\
    \hline
    Embodiment 
    & The level of corporeal presence of the metaverse content. 
    & The virtual content is only two-dimensional User Interface.
    & The virtual content is represented as 3D simple shapes and is static
    & The virtual content is dynamic (e.g., moving in the environment and performing animations).
    & The virtual content is not only represented as avatars or objects, but also have fixed effects on the virtual environment.
    & The virtual content has dynamic effects on the virtual environment such as particles effects.\\
    \hline
    
    Connectedness
    & The level of information communication between the metaverse and physical space (single and bi-directional communication).
    & There are no communications between objects.
    & Objects connect only on physical or virtual side. 
    & Objects could have single directional communication between virtual and physical. 
    & Objects could have bi-directional communications between virtual and physical through IoT broker. 
    & Objects have bi-directional communications with multiple IoT virtual and physical objects and multiple communication methods. 
    \\
    \hline
    
    Content Generation
    & The level of objects that are generated in the scene based on the context information. 
    & There are no generative content in the system. 
    & Context changes could affect the behaviours or shapes of virtual objects, but these are not generated in the scene. 
    & The shapes or behaviours of the objects are generated based on the context information through simple logic rules. 
    & The shapes and behaviours of the objects are generated by more complex logic states, such as L-systems and flocking. 
    & Shapes and behaviours of the objects are generated through advanced machine learning methods.
    \\
    \hline
    
    
    Mixed Reality Access (inspired by \cite{Milgram2011})
    & The level of mixed reality presence and immersion in the metaverse based on devices. 
    &Users could only access real environment or the virtual reality environment. 
    &Users could access low level of immersion with mixed reality such as using AR in mobile devices. 
    &Users could access mixed reality through head-mounted display devices (optical or video pass-through) with limited field of view. 
    &Users access mixed reality through head-mounted display devices (optical or video pass-through) with wide field of view. 
    &User can not only have high immersion in mixed reality, but also switch between virtual reality environment and real environment inside the devices. 
    \\
    \hline
    
    Context Awareness 
    &The level of context sensing capacity in the real environment. 
    &The system do not capture information from physical space.
    &The system could capture raw sensor data (as numeric values) from a single source.
    &The system could capture multiple raw sensor data through multiple sensor sources.
    
    &The system could capture, infer, and process complex data as input (such as real-time video object detection, or audio speech detection, resulting in a single classification (such as a number value or class label).
    
    &The system could understand complex data input, same as high, but resulting in more complex high-level or hierarchical context descriptions (such as detecting that a meeting is taking place, based on multiple sensor information).
    \\
    \hline
    
    Virtual-Physical Agency (inspired by \cite{Wooldridge1995}\cite{Holz2011})
    &The level of autonomous, reactive, proactive, and social capability of the entire metaverse ecosystem. 
    & The system does not have agent capabilities.
    & The system performs low levels of agent function in virtual or physical interactions, such as triggering a single event based on a sensor.
    & The system performs medium levels of agent function for either virtual or physical interactions, such as handling multiple events based on multiple sensors.
    &The agents should be capable of performing hybrid virtual and physical interactions and also has high agency (i.e., autonomous, reactive, proactive, and social interactions).
     &Same as high, but via more human-like, complex mechanisms and controllers, such as having cognitive or machine learning architectures.
    \\
    \hline
    
    User-Interactive Level (inspired by \cite{Zeltzer1992}\cite{Holz2011}) 
    &The level of control and feedback available for the user from the entire metaverse ecosystem. 
    
    & Users can not manipulate the simulated content.
    & Users could manipulate low degree of interaction with the simulated parameters of the virtual content, such as changing a Boolean value flag.
    & Users could manipulate medium degree of the simulated parameters, such as changing one or more values across a spectrum.
    & Users could manipulate high degree of the simulated parameters, such as manipulate multiple spectrum values, multiple objects, and multiple control types.
    & Same as high, but allow users to manipulate both physical and virtual objects to perform the interaction.
    \\
    \hline
    

  \end{tabular}
  \caption{Factors for comparison of the related prototypes (showed in Figure \ref{comparision}), and their meanings from very low to very high.}
  \label{Factors}
\end{table*}

\begin{figure}[tbh]
 \centering 
 \includegraphics[width=\linewidth]{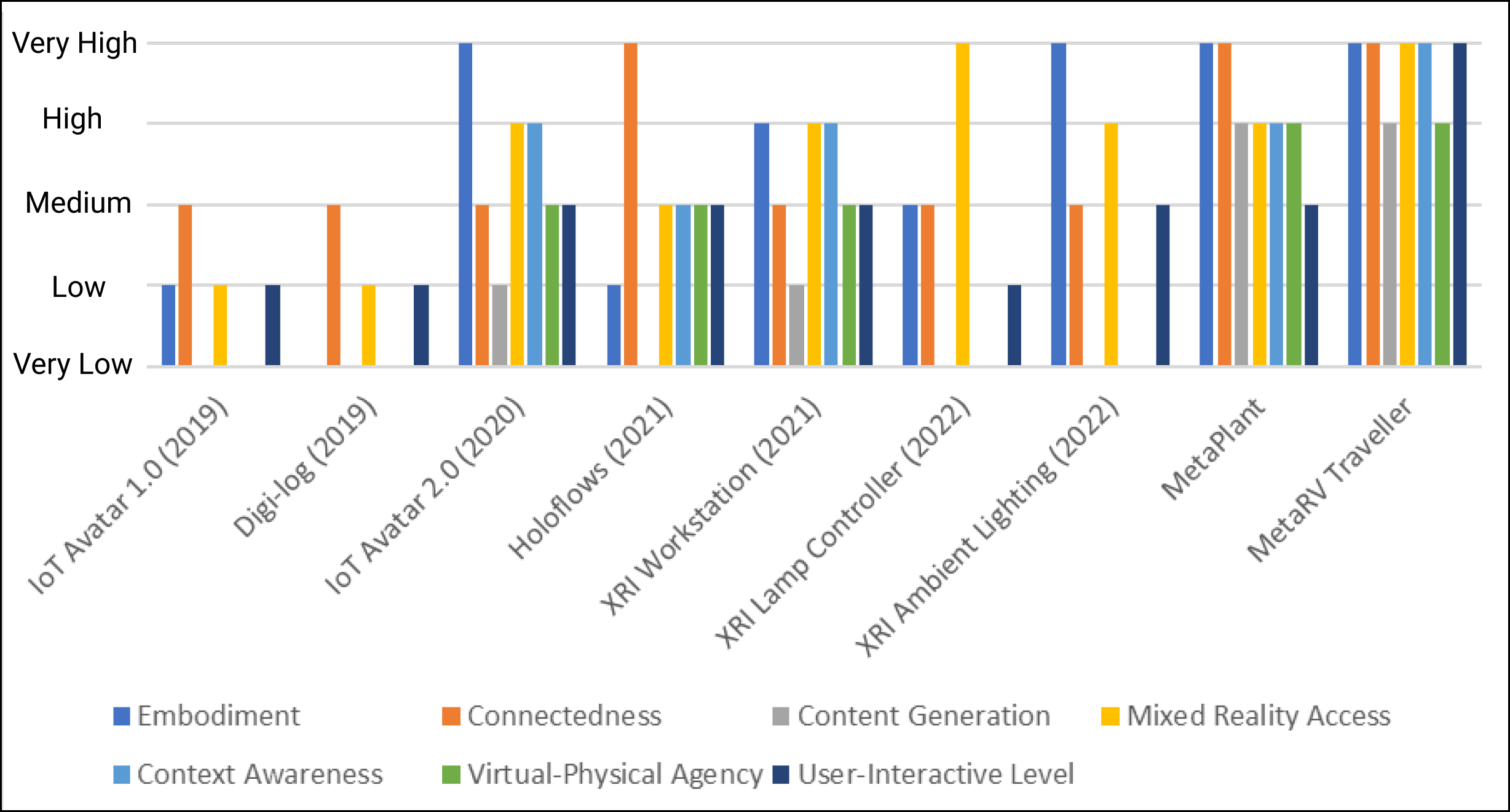}
 \caption{A comparison of the prototypes (based on the factors in Table \ref{Factors}) developed versus related work. It provides a subjective comparison with criteria in Table\ref{Factors} from 0-5 (very low to very high) of XR and IoT related projects, IoT Avatar 1.0\cite{shao2019iot}, IoT Avatar 2.0\cite{Morris2020IOTAvatar}, XRI Workstation\cite{Morris2021}, Digi-log \cite{jo2019Digi-log+}, Holoflows\cite{seiger2021holoflows}, XRI lamp Controller\cite{Guan2022IEEEVR}, XRI Ambient Lighting \cite{Guan2022IEEEVR}, and the two prototypes presented in this paper.}
 \label{comparision}
\end{figure}

Table \ref{Factors} presents the definitions of the factors for comparison, including embodiment, connectedness, content generation, mixed reality access, context awareness, virtual-physical agency and user-interactive level, with the meaning of the levels of very low, low, medium, high, and very high. Figure \ref{comparision} shows the authors' subjective comparison of levels of the related works with those of the prototypes presented in Section 4, MetaPlant and Meta-RV-Traveller, as described below. 
These show how the prototypes can be considered to address the merger of a broad range of needs and factors, at a high level. 

\subsection{Directions for Future Work}

The current work centered on the design and prototyping of these systems, and a more rigorous evaluation of these concepts is left as a direction for future research. For instance, more experimentation is needed to clearly show how cross-reality benefits users. Further, this work only considers a single-user cross-reality scenario, and multi-user explorations would be beneficial for proving the concept across different scenarios. New forms of representing changes in the physical and virtual environment and new design approaches are needed for users regardless of whether they are physically present, or virtually telepresent, across the entire cross-reality spectrum. Additionally, as the current work relies on subjective evaluation, a more thorough user evaluation is required to objectively assess the benefits and significance of this cross-reality research, as well as user-experience and performance. In the two example prototypes presented, the users can direct the virtual or physical world via their interactions, but the system-directed interactions to the user have not been examined, i.e., the cross-reality user interaction and feedback loop could be a significant factor for designers of future systems.

Further, more system performance and functionality evaluations are also required. It is also noted that new head-mounted displays and cross-reality hardware also remains to be incorporated into the framework, which would address several limitations related to visibility and immersion (i.e., use of color-passthrough HMD's versus the black \& white passthrough HMD used in this work). Lastly, an exploration of how to optimize the anchoring and placement of mixed reality objects without obstructing the user's field of view is also an important question for further research, such as positioning content in the peripheral field-of-view, or minimizing visual clutter.

\section{Summary}
This work has addressed the metaverse disconnect problem and the underlying challenges related to bridging and extending the metaverse concept across virtual, mixed, and physical environments through a single architectural framework. This framework has been instantiated to show two distinct prototype designs as cross-reality use cases wherein a user immersed in a mixed reality environment is able to engage with objects and elements from across either environment, and transition between these contexts while remaining in touch with each. This approach examines through design, outlining the need for connectivity, shared visualizations, and context awareness as ways to bridge the human-in-the-loop and the many virtual, mixed, and physical reality environments that they will increasingly engage in as the metaverse advances to maturity. Together, it is hoped that these will help foster the design-creation of a more seamless and extended metaverse environment.

\acknowledgments{
This work was supported by funding from the Tri-council
of Canada under the Canada Research Chairs program.}

\bibliographystyle{abbrv-doi}

\bibliography{template}
\end{document}